# The role of the coating and aggregation state in the interactions between iron oxide nanoparticles and 3T3 fibroblasts

**Malak Safi and Jean-François Berret***

Matière et Systèmes Complexes, UMR 7057 CNRS Université Denis Diderot Paris-VII
Bâtiment Condorcet, 10 rue Alice Domon et Léonie Duquet, 75205 Paris (France)

**Abstract.** Recent nanotoxicity studies revealed that the physico-chemical characteristics of engineered nanomaterials play an important role in the interactions with living cells. Here, we report on the toxicity and uptake of the iron oxide sub-10 nm nanoparticles by NIH/3T3 mouse fibroblasts. Coating strategies include low-molecular weight ligands (citric acid) and polymers (poly(acrylic acid), $M_W = 2000$ g mol$^{-1}$). We find that most particles were biocompatible, as exposed cells remained 100% viable relative to controls. The strong uptake shown by the citrate-coated particles is related to the destabilization of the dispersions in the cell culture medium and their sedimentation down to the cell membranes.

## 1. Introduction

Engineered nanoparticles are ultrafine colloids of nanometer dimensions with highly ordered crystallographic structures. These particles exhibit usually remarkable electronic, magnetic or optical properties that can be exploited in a variety of applications. In contrast to conventional chemicals however, the possible risks of using nanomaterials for human health and the environment have not been yet fully evaluated [1,2]. To estimate these risks, large research efforts were directed towards the development of toxicology assays. The objectives of these assays are the quantitative determination of the viability of living cells that were incubated with nanomaterials [1,2]. In recent years, the viability of many different cell lines was investigated with respect to a wide variety of engineered nanoparticles. Recent reviews attempted to recapitulate the main features of the toxicity induced by nanomaterials. One of these features deals with the coating of the particles. In most *in vitro* studies, the chemistry of the interfaces between the nanocrystals and the solvent was anticipated to be a key feature of cell-nanoparticle interactions.

In the present paper, we investigated the *in vitro* toxicity and internalization of sub-10 nm iron oxide (maghemite, $\gamma$-Fe$_2$O$_3$) nanoparticles using mice

* Corresponding author. Tel.: +33 1 5727 6147; fax: +33 1 5727 6211.
*E-mail address*: jean-francois.berret@univ-paris-diderot.fr .

NIH/3T3 fibroblasts. We have also developed an easy and widely applicable method to adsorb ion-containing polymers onto the nanoparticle surfaces [3,4]. We have found that this low-molecular weight polymers augmented the hydrodynamic diameters of the particles by only 4 nm, and at the same time preserved the long term colloidal stability in most water based solvents, including buffers and cell culture media [5]. This noticeable increase in stability as compared to classical ligand-coated particles has prompted us to perform toxicity assays, and to explore the effect of the dispersion state on intracellular uptake.

## 2. Experimental/Methodology

The synthesis of iron oxide nanoparticles used the technique of « soft chemistry » based on the polycondensation of metallic salts in alkaline aqueous media. The synthesis has been previously described, and we refer to this work for more details [6]. In the present study, two maghemite batches were synthesized with median diameters in the range of 7 - 8 nm and a narrow polydispersity. In this work, two different coating agents were utilized : citric acid molecules and poly(acrylic acid) polymers. For citric acid, the complexation of the surface charges was performed during the synthesis by adding tri-sodium citrate in excess under vigorous stirring, followed by washing steps with acetone. To adsorb polyelectrolytes on the surface of the nanoparticles, we followed the precipitation-redispersion protocol [3,4]. The precipitation of iron oxide dispersions by $PAA_{2K}$ was performed in acidic conditions (pH 2). The precipitate was separated from the solution by centrifugation, and its pH was increased by addition of ammonium hydroxide. The precipitate redispersed spontaneously at pH 7 - 8, yielding a clear solution that contained the polymer coated particles.

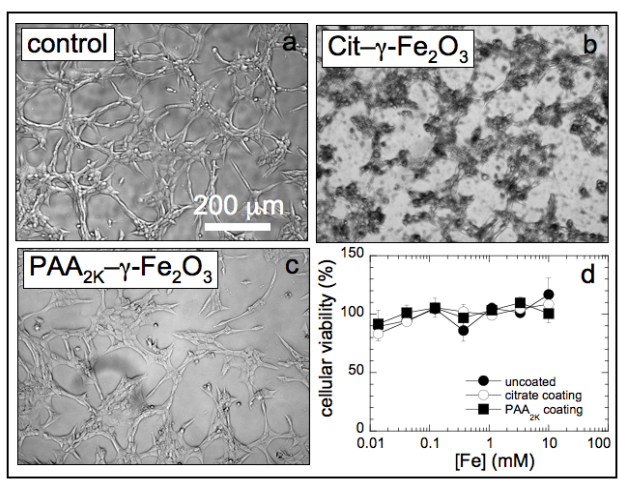

**Figure 1** : Transmission optical microscopy (10×) images of NIH/3T3 fibroblasts (a). In (b) and (c), the cells were incubated with Cit–γ-$Fe_2O_3$ and $PAA_{2K}$–γ-$Fe_2O_3$ during 24 h at a concentration of 1 mM. (d) MTT viability assays conducted on NIH/3T3 cells incubated with bare, citrate and $PAA_{2K}$ coated particles.

NIH/3T3 fibroblast cells from mice were grown as a monolayer in Dulbecco's Modified Eagle's Medium (DMEM) with High Glucose (4.5 g $L^{-1}$) and stable Glutamine (PAA Laboratories GmbH, Austria). This medium was supplemented with 10% Fetal Bovine Serum (FBS), and 1% penicillin/streptomycin (PAA Laboratories GmbH, Austria). Exponentially growing cultures were maintained in a humidified atmosphere of 5% $CO_2$ - 95% air at 37°C, and under these conditions the plating efficiency was 70 – 90 % and the doubling time was 12 – 14 h. MTT assays were performed with coated and uncoated iron oxide nanoparticles for metal molar concentrations [Fe] between 10 µM to 10 mM. Cells were seeded into 96-well microplates, and the plates were placed in an incubator overnight to allow for attachment and recovery.

## 3. Results and Discussion

*Optical microscopy* : In order to determine their optimal growth conditions, the fibroblasts were first plated in culture medium without particles. Fig. 1a provides an illustration of the NIH/3T3 observed by optical microscopy at a 50 % coverage (objective 10×). Figs. 1b and 1c display NIH/3T3 fibroblasts that were exposed during 24 h with Cit–γ-$Fe_2O_3$ and $PAA_{2K}$–γ-$Fe_2O_3$ nanoparticles respectively, at [Fe] = 1 mM. Note that for contrast reasons the supernatant containing the citrate-coated particles was removed and after thorough washing with PBS, it was replaced by pristine medium. For the $PAA_{2K}$-coated particles, the images were recorded in the same conditions as for the control, the particles being dispersed in the cell medium. In Fig. 1, there is a marked difference between cells incubated with citrate and with $PAA_{2K}$-coated particles. Due to a massive internalization and/or adsorption of the nanomaterial by the cells, the cells exposed to the citrate-coated particles were more difficult to detect. The dark patterns seen in the bottom left image were stemming from internalized/adsorbed Cit–γ-$Fe_2O_3$. In contrast, the cells incubated with $PAA_{2K}$–γ-$Fe_2O_3$ behaved as the control.

*MTT assays* : Toxicity assays alone can quantify the cell viability under nanoparticles exposure. MTT (3-(4,5-dimethylthiazol-2-yl)-2,5-diphenyl tetra-zolium bromide) viability assays were conducted on NIH/3T3 cells for molar concentrations [Fe] varying from 10 µM to 10 mM and incubation time of 24 h. Fig. 1d displays the cellular viability as a function of the molar concentrations [Fe]. Uncoated, citrate-coated and $PAA_{2K}$-coated particles were surveyed. For the three iron oxide samples, the viability remained at a 100 % level within the experimental accuracy. These findings indicate a normal mitochondrial activity for the cultures. Our results are in good agreement with earlier reports from the literature which showed that both crystalline forms of iron oxide, maghemite $\gamma$-$Fe_2O_3$ and magnetite $Fe_3O_4$ were found biocompatible with respect to cell cultures [1].

*Nanoparticle uptake* : The uptake of nanoparticles by the cells was monitored by UV-Visible spectrometry. Aliquots of the supernatants in contact with the cells were collected at regular time intervals and analyzed with respect to their oxide concentration. In the following, we assume that the nanoparticles not present in the supernatant, and thus not detected by spectrometry were either adsorbed on the cellular membranes or taken up by the cells [7]. This assumption allowed us to evaluate the mass of metallic atoms $M_{Fe}$ internalized or adsorbed by the cells as a function of the time. Thanks to proliferation (data not shown), $M_{Fe}$ was normalized with respect to the number of cells present at a given time, resulting in masses expressed in picogram of iron per cell. Fig. 2 compares the mass of iron $M_{Fe}$ incorporated or adsorbed per cell for Cit–$\gamma$-$Fe_2O_3$ and $PAA_{2K}$–$\gamma$-$Fe_2O_3$. $M_{Fe}$ was found to be unchanged for the polymer coated particles whereas it increased logarithmically with time for the citrate-coated particles. After a 24 h incubation, the mass per cell reached a value of 250 pg, a result that compared well to those of the literature on human fibroblasts [7].

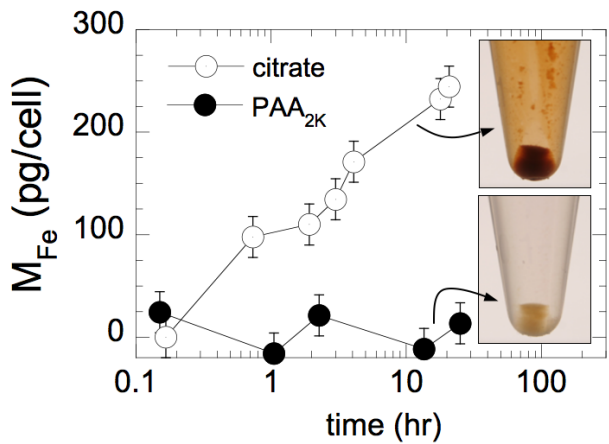

*Figure 2* : Mass $M_{Fe}$ of iron incorporated by the NIH/3T3 cells for experiments conducted with Cit–$\gamma$-$Fe_2O_3$ and $PAA_{2K}$–$\gamma$-$Fe_2O_3$. Inset : Centrifugation pellets of NIH/3T3 cells that were grown with citrate and $PAA_{2K}$-coated particles.

The inset of Fig. 2 displays centrifugation pellets of NIH/3T3 cells that were incubated with Cit–$\gamma$-$Fe_2O_3$ and $PAA_{2K}$–$\gamma$-$Fe_2O_3$ nanoparticles for 24 h. After incubation, the cultures were washed thoroughly using PBS in order to remove particles loosely adsorbed onto the cellular membranes. The surface of the flasks were then mechanically scraped and the cell suspensions were centrifuged in Eppendorf vials. The exposure of cells to Cit–$\gamma$-$Fe_2O_3$ resulted in the significant darkening of the centrifugation pellet as compared to the $PAA_{2K}$–$\gamma$-$Fe_2O_3$ treated cell line (inset of Fig. 2).

It may seem surprising that the NIH/3T3 cells behaved differently with respect to particles which have the same charges at their surfaces. Cell membranes are indeed known to be negatively charged in average and therefore these membranes should exert a net electrostatic repulsion towards surrounding diffusing particles of the same charges. To understand the differences in uptake between citrate and $PAA_{2K}$-coated particles, the colloidal stability of the particles in various solvents, including brines, buffers and cellular growth media was recently put under scrutiny [5]. Here, we underscore the results obtained when the particles were dispersed in the complete culture medium. In DMEM, we found that the citrate-coated nanoparticle precipitate instantaneously, whereas the $PAA_{2K}$-coated particles remained dispersed (*i.e.* unaggregated) over weeks. We anticipate that the pronounced uptake exhibited by Cit–$\gamma$-$Fe_2O_3$ is related to the destabilization of the initially dispersed nanoparticles and their accumulation by gravity in the vicinity of the cell membranes [2].

## 4. Conclusion

In this work, the toxicity and uptake of iron oxide nanoparticles by NIH/3T3 fibroblasts were investigated. The proliferative properties of the cells and their viability in presence of engineered nanomaterials were evaluated by *i)* transmission optical microscopy to determine the growth laws of the cell populations, *ii)* MTT assays as a function of the metal dose and *iii)* UV-Visible spectrometry for the estimation of the particles uptake. The stronger uptake shown by Cit–$\gamma$-$Fe_2O_3$ was related to the destabilization of the initially dispersed nanoparticles in the cell culture medium and their sedimentation near by the surfaces of the cells. For the $PAA_{2K}$-coated particles, the polymer coating ensured a long term (> year) stability even in physiological conditions. For these particles, the uptake resulted only by diffusion and single adsorption on the cell membranes.